\documentclass[sigconf]{acmart}
\AtBeginDocument{%
  }

\copyrightyear{2026}
\acmYear{2026}
\setcopyright{cc}
\setcctype{by}
\acmConference[MM '26]{Proceedings of the 34th ACM International Conference on Multimedia}{November 10--14, 2026}{Rio de Janeiro, Brazil}
\acmBooktitle{Proceedings of the 34th ACM International Conference on Multimedia (MM '26), November 10--14, 2026, Rio de Janeiro, Brazil}
\acmDOI{10.1145/3767308.3834962}
\acmISBN{979-8-4007-2213-4/2026/11}

\begin{document}
\title[SubtleTalk: Generating Controllable Weakly-correlated Facial Dynamics for 3D Talking Heads \\ via Residual Flow Matching]{SubtleTalk: Generating Controllable Weakly-correlated Facial Dynamics for 3D Talking Heads via Residual Flow Matching}

\author{Chenyang Ding}
\orcid{0009-0008-6079-0688}
\affiliation{%
  \institution{Shanghai Jiao Tong University}
  \city{Shanghai}
  \country{China}
}
\email{chenyangding@sjtu.edu.cn}

\author{Shuai Tan}
\orcid{0000-0003-3322-5161}
\affiliation{%
  \institution{Shanghai Jiao Tong University}
  \city{Shanghai}
  \country{China}
}
\email{tanshuai0219@sjtu.edu.cn}

\author{Qunfen Lin}
\orcid{0009-0002-5791-5108}
\affiliation{%
  \institution{Tencent Games}
  \city{Shenzhen}
  \country{China}
}
\email{volleylin@tencent.com}

\author{Xinwei Jiang}
\orcid{0000-0002-1766-3332}
\affiliation{%
  \institution{Tencent Games}
  \city{Shenzhen}
  \country{China}
}
\email{wesleyjiang@tencent.com}

\author{Zijiao Zeng}
\orcid{0009-0008-0378-373X}
\affiliation{%
  \institution{Tencent Games}
  \city{Shenzhen}
  \country{China}
}
\email{zijiaozeng@tencent.com}

\author{Ye Pan}
\correspondingauthor
\orcid{0000-0003-0355-989X}
\affiliation{%
  \institution{Shanghai Jiao Tong University}
  \city{Shanghai}
  \country{China}
}
\email{whitneypanye@sjtu.edu.cn}

\renewcommand{\shortauthors}{Chenyang Ding et al.}
\begin{abstract}
Audio-driven 3D facial animation aims to synthesize realistic and temporally coherent motions from speech. Despite notable progress in lip synchronization, weakly correlated dynamics, including eyebrow movements, eye blinks, and head motion, which are essential to photorealistic facial animation, remain difficult to model faithfully and often appear static or unnaturally repetitive. We attribute this limitation to three factors: (a) insufficient conditioning for weakly correlated dynamics; (b) the limited ability of deterministic regression to capture diverse motion patterns; (c) data bottlenecks from unreliable upper-face pseudo-labels and limited dataset diversity. To address these issues, we propose \textbf{SubtleTalk}, a framework for generating natural and controllable weakly correlated facial dynamics via multi-condition modeling and residual flow matching. First, to compensate for the limited guidance of speech alone, we introduce interpretable controls, including prosody, regional intensity, and Valence-Arousal (VA) signals, to explicitly capture the timing, magnitude, and affective variation of weakly correlated dynamics. Second, to overcome the limited expressiveness of deterministic regression, we build residual flow matching based on a stable speech-driven motion prior, allowing the model to capture stochastic deviations beyond deterministic prediction. Third, to alleviate the data bottleneck, we construct \textbf{SubtleTalk-Face}, a large-scale 3D facial animation dataset comprising about 3,900 identities and 74 hours of data, built via a simple and scalable pseudo-labeling pipeline and featuring improved upper-face tracking and frame-level VA annotations. Extensive experiments demonstrate that our method significantly improves the realism and diversity of weakly correlated facial dynamics while preserving accurate lip synchronization. Our project page is available at \url{https://molly-ding.github.io/SubtleTalk/}.
\end{abstract}



\begin{teaserfigure}
  \begin{center}
    \includegraphics[width=0.9\textwidth]{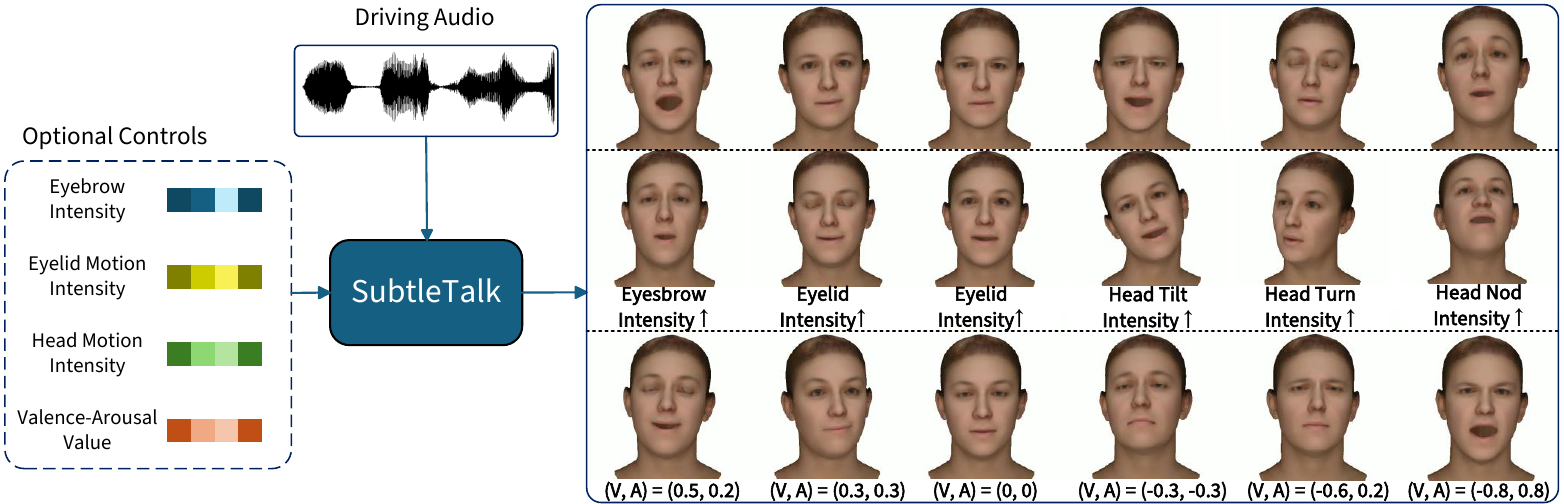}
  \end{center}
  \caption{\textbf{Examples generated by SubtleTalk.} Given driving audio and optional controls for regional intensity and Valence-Arousal (VA), SubtleTalk can synthesize natural 3D facial animations. It enables explicit control over eyebrow , eyelid , and head dynamics, and defaults to expressive audio-only generation when auxiliary controls are omitted.}
  \Description{Overview of SubtleTalk.}
  \label{fig:teaser}
\end{teaserfigure}

\maketitle

\section{Introduction}
Audio-driven 3D facial animation aims to synthesize realistic 3D facial motions from speech~\cite{tan2023emmn, tan2024style2talker, tan2024flowvqtalker, tan2024say, tan2025fixtalk, tan2025edtalk++, tan2024animate, tan2025animate, tan2026synmotion, tan2026codance,ji2025sport,ma2026phys,ji2025pomp}. 
Producing photorealistic and expressive facial animation requires not only accurate lip synchronization but also realistic eyebrow movements, eye blinks, and head motion. Although prior works have achieved strong lip-sync performance~\cite{VOCA, MeshTalk, FaceFormer, CodeTalker, EmoTalk, EMOTE, SelfTalk, FaceDiffuser}, eyebrow movements, eye blinks, and head motion that are weakly correlated with speech remain difficult to model faithfully. 
Specifically, eyebrow movements are often oversmoothed into static, rigid~\cite{MeshTalk, FaceFormer, CodeTalker, FaceDiffuser, DiffPoseTalk} or speech-irrelevant repetitive patterns~\cite{ProbTalk3D, ARTalk}, while eye blinks may be missing altogether~\cite{FaceFormer, CodeTalker} or reduced to incomplete closures~\cite{EmoTalk, FaceDiffuser, ProbTalk3D, DiffPoseTalk, ARTalk}. Moreover, although recent methods~\cite{DiffPoseTalk, MMHead, ARTalk} can generate basic head movements, the resulting motion amplitudes are typically confined to an average range, difficult to model the diverse range of exaggerated or subtle head motions needed for varied scenarios.

We attribute these limitations to three primary factors: (1) Insufficient conditioning. Speech provides only weak guidance for upper-face and head dynamics, especially sparse, transient events such as eyebrow movements and eye blinks. As a result, models tend to regress toward averaged patterns, yielding over-smoothed predictions and weakened subtle facial cues. (2) Limited diversity modeling. Deterministic regression methods~\cite{VOCA, FaceFormer, EmoTalk, EMOTE} struggle to model the one-to-many mapping from speech to facial motion, as their objectives bias predictions toward conditional averages. Discrete-prior methods~\cite{MeshTalk, CodeTalker, MMHead, ProbTalk3D, ARTalk} partly mitigate this issue, but quantization error and codebook instability still limit fine-grained fidelity. (3) Data bottlenecks. Weakly correlated dynamics require large-scale, high-quality, in-the-wild data, but controlled 3D capture~\cite{BIWI, VOCA, MeshTalk, Multiface} is difficult to scale to natural scenes. Consequently, recent datasets~\cite{EmoTalk, MEDTalk, EMOTE, MMHead, DiffPoseTalk} mainly rely on monocular 3D reconstruction~\cite{SPECTRE, EMOCA, MICA, LiveLink, MediaPipe} from videos. However, existing FLAME-based datasets~\cite{EMOTE, MMHead, DiffPoseTalk} often use early fitting pipelines~\cite{SPECTRE, EMOCA, MICA}, resulting in weak upper-face fidelity.

To address these issues, we propose \textbf{SubtleTalk}, which generates natural weakly correlated facial dynamics from speech and optional control signals while preserving accurate lip synchronization, as shown in Figure~\ref{fig:teaser}. Our key insight is threefold: (1) augment speech with explicit signals for event timing, motion magnitude, and fine-grained affective state; (2) separate strongly speech-correlated motion from ambiguous upper-face and head dynamics, preserving accurate lip synchronization while modeling residual uncertainty via residual flow matching; and (3) support learning of such subtle dynamics with large-scale, high-fidelity supervision, especially for upper-face motion. 

First, we model event timing with multi-scale WavLM~\cite{WavLM} features and explicit prosodic cues, including fundamental frequency ($F_0$) and energy ($E$), inspired by prior findings that brow and head movements are aligned with prosodic structure and pitch accents \cite{ProsodyInfo, MotionWithProsody1, MotionWithProsody2}. We represent motion magnitude with disentangled regional intensities, and use continuous frame-level Valence-Arousal (VA)~\cite{VA} signals to provide fine-grained affective cues, avoiding the coarse temporal control of global labels and the limited granularity of discrete categories in capturing subtle affective variation. Second, we adopt a predict-and-refine residual flow-matching framework that separates strongly speech-correlated motion from ambiguous upper-face and head dynamics to concentrate generative capacity on the more ambiguous components. 
Third, we construct \textbf{SubtleTalk-Face} from large-scale, carefully filtered in-the-wild videos with a stronger 3D reconstruction pipeline. This provides broader, more faithful supervision, especially for upper-face motion, and better supports learning natural weakly correlated dynamics. 

Our contributions are summarized as follows:
\begin{itemize}
    \item We propose SubtleTalk, a two-stage residual flow-matching framework for natural and controllable weakly correlated 3D facial dynamics with accurate lip synchronization.
    \item We introduce an explicit multi-condition modeling strategy that combines regional intensities, VA signals, and speech prosody to mitigate over-smoothing in transient dynamics.
    \item We construct SubtleTalk-Face, a large-scale 3D facial animation dataset with 3,900 identities and 74 hours of data, featuring high-fidelity upper-face tracking and VA labels.
    \item Extensive experiments show that SubtleTalk outperforms existing methods in synthesizing realistic weakly correlated facial dynamics.
\end{itemize}
\section{Related Work}
\subsection{Audio-driven 3D Facial Animation}
Audio-driven 3D facial animation aims to synthesize realistic facial motions from speech. Early learning-based methods typically formulated this as a deterministic regression from audio to 3D vertex trajectories or facial parameters \cite{VOCA, FaceFormer, EmoTalk, EMOTE, SelfTalk, UniTalker}.
For example, SelfTalk~\cite{SelfTalk} improves lip synchronization through a self-supervised framework jointly constrained by speech recognition and lip-reading consistency. UniTalker~\cite{UniTalker}, in contrast, trains a unified model on multiple datasets with heterogeneous 3D annotations using shared representations and separate mapping heads, thereby improving lip-sync accuracy. However, these methods still optimize motion prediction mainly with reconstruction-based regression losses, which tend to yield conditional-average solutions for under-constrained dynamics, particularly subtle eyebrow motion, eye blinks, and head movements.

To enhance generation diversity, some works introduce discrete motion priors into facial motion space \cite{MeshTalk, CodeTalker, ProbFace, ProbTalk3D, MMHead, LSF-Animation, DEEPTalk, ARTalk, MemoryTalker}. A representative work is CodeTalker~\cite{CodeTalker}, which introduces a discrete motion prior based on VQ-VAE \cite{VQVAE} for autoregressive motion generation. Building on this paradigm, ProbFace~\cite{ProbFace}, DEEPTalk~\cite{DEEPTalk}, and ARTalk~\cite{ARTalk} further combine quantized motion representations with hierarchical temporal modeling to better model the stochastic and one-to-many mapping from speech to facial motion. Although these methods partially alleviate mean-collapse, discretizing continuous facial motion space inevitably introduces reconstruction error and may still suffer from codebook collapse, training instability, and repetitive artifacts.

More recently, diffusion-based methods have been introduced to model facial motion directly in continuous spaces~\cite{FaceDiffuser, DiffSpeaker, Media2Face, FaceTalk, DiffPoseTalk, ProsodyTalker, THUNDER, Audio2Face3D}. Although DiffPoseTalk~\cite{DiffPoseTalk} enhances expressiveness through joint facial and head motion generation, it still struggles to faithfully model dynamics that are weakly correlated with speech. Specifically, fine-grained eyebrow motions and complete eye closures are often inadequately represented, while head movements are restricted to average ranges rather than providing the diverse subtle or exaggerated motions needed for varied scenarios. Meanwhile, diffusion-based formulations still require costly iterative denoising during inference, which motivates Flow Matching~\cite{FlowMatching} as a more efficient continuous alternative in motion synthesis~\cite{MotionFlowMatching}.

\subsection{Conditional 3D Talking Head Generation}
Speech alone provides limited guidance for continuously varying emotional states and the motion magnitude of weakly correlated transient events. To reduce this ambiguity, some methods introduce auxiliary controls based on speaker style, either through temporally invariant identity labels~\cite{VOCA, FaceFormer, CodeTalker} or style features extracted from reference motions~\cite{DiffPoseTalk, ARTalk}. Although these strategies can enrich stylistic diversity, they often limit flexibility by relying on seen training identities or requiring cumbersome reference-based style extraction. Moreover, because such style representations are implicit and weakly interpretable, they do not readily support direct, fine-grained control over localized motion amplitudes.

To further improve expressiveness, some works~\cite{EMOTE, EmoTalk, EmoFace} employ discrete emotion labels, while DEITalk~\cite{DEITalk} introduces dynamic emotion intensities derived from a speech-expression space. More recently, multimodal methods~\cite{Media2Face, AVI-Talking, ExpCLIP, MEDTalk} explore visual or text prompts for personalized styles. However, existing approaches still lack a clear and continuous control space for fine-grained emotional dynamics. In contrast, our method adopts a more interpretable conditioning design by explicitly incorporating continuous frame-level VA signals, regional motion intensities, and prosody-aware audio features, which enables fine-grained control over localized motion amplitudes and emotional variation.

\subsection{3D Facial Animation Datasets}
3D facial animation datasets are fundamental for developing natural talking head synthesis models. Early controlled laboratory captures~\cite{BIWI, VOCA, MeshTalk, Multiface} provide accurate ground truth but suffer from limited scale, identity diversity, and applicability to in-the-wild scenarios. To alleviate this bottleneck, subsequent works~\cite{EmoTalk, EmoFace, EMOTE, MMHead, DiffPoseTalk, MEDTalk} attempt to construct larger datasets by fitting 3D parametric models~\cite{FLAME, Blendshape, MetaHuman} to 2D portrait videos via monocular reconstruction pipelines~\cite{SPECTRE, EMOCA, MICA, LiveLink, MediaPipe}.

We adopt the FLAME representation to facilitate standard evaluation. However, existing FLAME-based datasets~\cite{EMOTE, MMHead, DiffPoseTalk} rely on early fitting algorithms~\cite{SPECTRE, EMOCA, MICA} with limited upper-face fidelity~\cite{REALY}. Moreover, we find that even large-scale datasets like MMHead~\cite{MMHead} lack sufficient visual cleaning for extreme poses or occlusions, which severely degrades pseudo-label quality. To address this, we introduce \textbf{SubtleTalk-Face}, a high-quality, 73.8-hour dataset of roughly 3,905 identities curated from multiple collections~\cite{VFHQ, HDTF, CelebV-HQ, MEAD, DiffPoseTalk}. After applying strict filtering criteria to eliminate poor lip-sync, extreme poses, and occlusions, \textbf{SubtleTalk-Face} integrates synchronized audio-visual data with continuous VA annotations~\cite{EmotiEffLib}. Furthermore, it employs an advanced reconstructor~\cite{TEASER} to extract high-fidelity FLAME pseudo-labels for finer localized facial dynamics.


\begin{figure*}[t]
  \centering
  \includegraphics[width=0.835\textwidth]{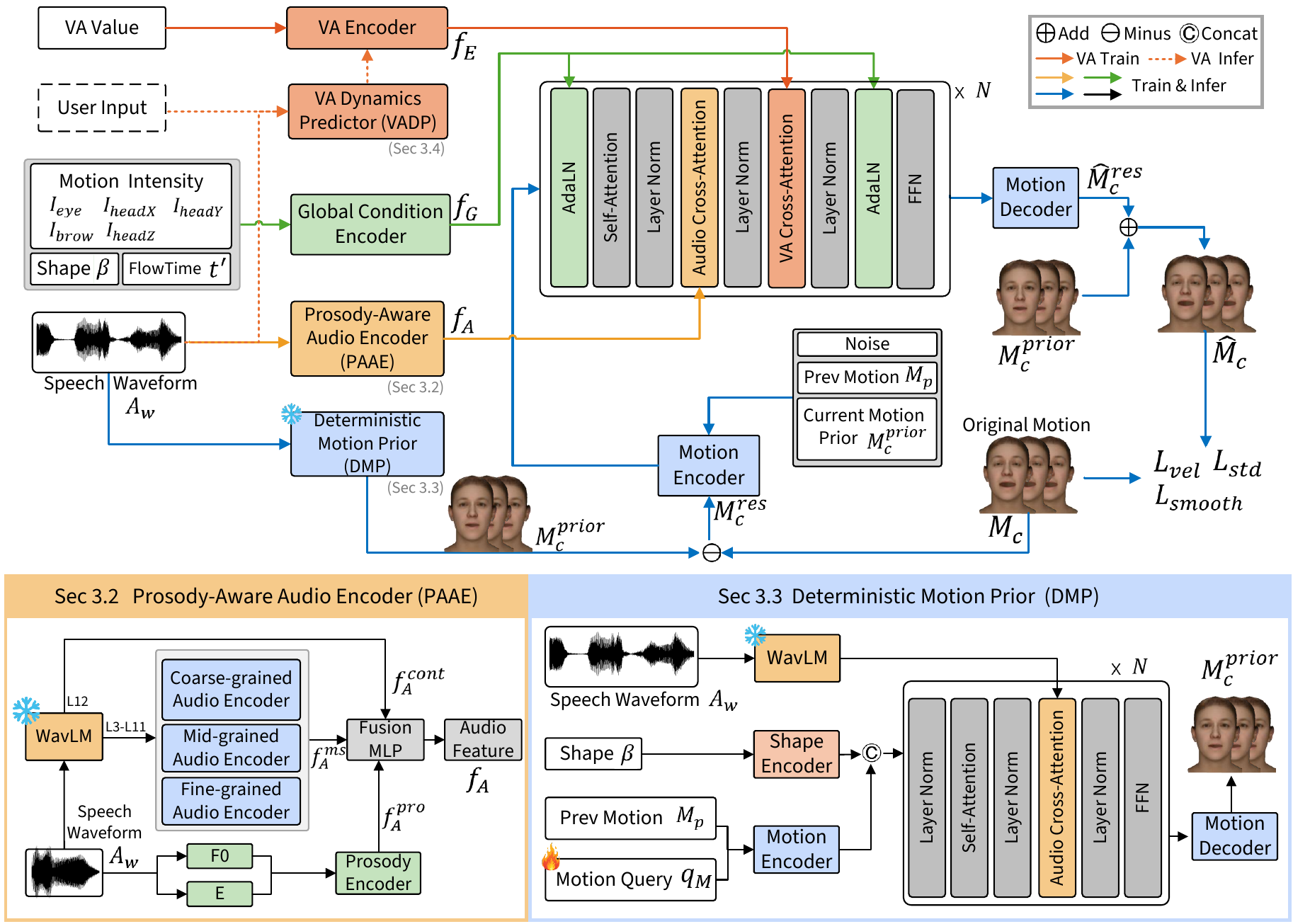}
\caption{
Overview of SubtleTalk, a two-stage framework for controllable weakly correlated facial dynamics. A deterministic prior (DMP) first anchors speech-driven lip motion, while residual flow matching model captures diverse upper-face variations. Multimodal conditioning (prosody, regional intensity, and VA) provides explicit control, and 
VA Dynamics Predictor (VADP) supports fully automatic or user-guided synthesis.
}
  \Description{Pipeline of SubtleTalk(img1).}
  \label{fig:pipeline1}
\end{figure*}

\section{Method}
Figure~\ref{fig:pipeline1} illustrates the overall framework of SubtleTalk. Given a speech window $A_w$, we generate the target motion segment $M_c$ conditioned on $A_w$ and the previous motion context $M_p$. We first construct explicit multimodal conditions (Sec.~\ref{sec_3_2}), including prosody-aware enhanced audio features, regional intensity controls, and continuous VA signals, to better constrain weakly correlated dynamics. We then adopt a two-stage architecture (Sec.~\ref{sec_3_3}), where the Deterministic Motion Prior (DMP) predicts a stable lip-synchronized prior $M_c^{\mathrm{prior}}$, and the residual flow matching stage models stochastic variations beyond it. Finally, we introduce a robust and adaptive control design (Sec.~\ref{sec_3_4}), where Implicit Disentangled Control (IDC) improves control fidelity by reducing cross-factor interference, and the VA Dynamics Predictor (VADP) enables both automatic and user-controllable inference.
\subsection{Preliminaries}
\label{sec_3_1}
\paragraph{\textbf{3D Face Representation.}}
We employ FLAME~\cite{FLAME} to represent 3D facial geometry using a set of parameters $\{\beta, \psi, \theta\}$. Following the setup in the 3D face reconstruction works~\cite{TEASER}, $\beta \in \mathbb{R}^{300}$ is the identity-specific shape parameters, $\psi \in \mathbb{R}^{50}$ is the expression parameters, and $\theta \in \mathbb{R}^6$ is the pose parameters, including 3D head rotation and jaw opening. Given these parameters, the 3D facial mesh $\mathbf{V} \in \mathbb{R}^{N \times 3}$ ($N=5023$ vertices) can be obtained via $\mathbf{V} = \text{LBS}(\beta, \psi, \theta)$, where $\text{LBS}(\cdot)$ denotes the Linear Blend Skinning function.

\paragraph{\textbf{Flow Matching.}}
Flow Matching~\cite{FlowMatching} learns a conditional velocity field that transports a simple source distribution to the target data distribution. Given a target sample $x_1 \sim q(x_1)$ and a source sample $x_0 \sim p_0=\mathcal{N}(0,\mathbf{I})$, the standard formulation defines the interpolation path as $x_{t'}=(1-t')x_0+t'x_1$, where $t'\sim\mathcal{U}(0,1)$, with target velocity $u_{t'}=\frac{d x_{t'}}{dt'}=x_1-x_0$. Conditioned on $c$, the network $V_\theta$ predicts the conditional velocity field $V_\theta(x_{t'},t',c)$ and is optimized by $\mathcal{L}_{\mathrm{FM}}=\mathbb{E}_{x_0,x_1,t'}\left[\|V_\theta(x_{t'},t',c)-(x_1-x_0)\|_2^2\right]$.

\paragraph{\textbf{Problem Formulation.}}
Given an audio sequence $A_{total}$, the goal is to generate the aligned facial motion sequence $M_{total}=\{m_t\}_{t=1}^{T_{total}}$, where $m_t=\{\psi_t,\theta_t\}$ denotes the motion at frame $t$. To model temporal dependencies, we divide the full sequences into fixed-length local windows $A_w=[A_p;A_c]$ and $M_w=[M_p;M_c]$, where $T_w=T_p+T_c$, with $T_p$ previous-context frames and $T_c$ current frames to predict. During training, the model predicts $M_c$ conditioned on $(A_p,M_p,A_c)$. Under the flow matching formulation, we set $x_1=M_c$ and learn a conditional velocity field that transports samples from a simple prior to the target segment. At inference, arbitrary-length sequences are processed with a sliding-window strategy: for the $k$-th window, the model predicts $M_c^k$ from $(A_p^k,M_p^k,A_c^k)$, then uses $M_c^k$ as $M_p^{k+1}$ for the next window. The full motion sequence $M_{total}$ is obtained by concatenating all predicted segments with boundary smoothing.

\subsection{Explicit Driving Signals}
\label{sec_3_2}
We introduce three complementary driving signals: prosody-aware audio cues, regional intensity signals, and continuous VA signals to explicitly model the timing, magnitude, and affect of weakly correlated facial dynamics.

\paragraph{\textbf{Prosody-Aware Audio Encoder (PAAE).}}
As shown in Figure~\ref{fig:pipeline1}, we organize speech conditioning into three branches: a content branch, a multi-scale acoustic branch, and a prosody branch. This design is motivated by prior analyses showing that higher self-supervised speech layers are more content-oriented, while lower and intermediate layers retain richer prosodic information~\cite{WavLM, SUPERB_Prosody}. Given a speech window $A_w$, we feed it into a frozen \emph{WavLM-base}~\cite{WavLM} with 12 Transformer layers and use the temporally aligned last-layer feature as the content feature $f_A^{cont}$. For multi-scale acoustic modeling, we take intermediate hidden states $\{H^{(l)}\}_{l=3}^{11}$, where $l$ denotes the layer index, aggregate them with learnable layer weights, and pass them through fine-, mid-, and coarse-grained temporal convolution branches, whose outputs are adaptively fused into a multi-scale acoustic feature $f_A^{ms}$. In parallel, we extract frame-level $F0$ and log-energy $E$ from the waveform and encode them with a lightweight prosody encoder to obtain a prosody feature $f_A^{pro}$. Finally, a lightweight fusion MLP integrates $f_A^{cont}$, $f_A^{ms}$, and $f_A^{pro}$ into a unified prosody-aware enhanced audio feature $f_A$.

\paragraph{\textbf{Regional Intensity Signals.}}
Inspired by prior works on explicit intensity control~\cite{FantasyTalking} and region-wise facial manipulation~\cite{EDTalk}, we define region-wise intensity signals directly in the target 3D motion space. For each current motion window $M_c$, we compute five scalar controls $I=[I_{eye}, I_{brow}, I_{headX}, I_{headY}, I_{headZ}]$, corresponding to eye, brow, and head nod/turn/tilt intensity, respectively. 
For the eye and brow regions, intensity is defined as the temporal standard deviation of the corresponding FLAME vertex trajectories within the window. 
For head motion, $I_{headX}$, $I_{headY}$, and $I_{headZ}$ are defined as the temporal standard deviations of the corresponding FLAME head-pose parameters $\theta_X$, $\theta_Y$, and $\theta_Z$. 
After normalization with dataset statistics, these scalars are concatenated into an intensity vector, which is then fed into the Global Condition Encoder for subsequent conditioning.

\paragraph{\textbf{Valence-Arousal Signals.}}
As discussed in Sec.~1, we adopt continuous VA signals as a compact yet fine-grained affect representation. Specifically, we extract frame-wise visual VA labels using a pretrained affect estimator~\cite{EmotiEffLib} and use the resulting sequence as the affective control signal during training. To better model the affect-to-motion relationship, we apply Fourier feature~\cite{FourierFeatures} encoding to VA before projection, since VA serves as a higher-level affect cue rather than an explicit motion control signal, and its low-dimensional continuous form makes it difficult for MLPs to capture rich frequency variations
Specifically, for each frame, given $e_t=[v_t,a_t]\in[-1,1]^2$, we encode it as
\begin{equation}
\mathrm{FE}(e_t)=\big[e_t,\{\sin(2^k\pi e_t),\cos(2^k\pi e_t)\}_{k=0}^{K-1}\big],
\end{equation}
where $K$ is the number of frequency bands. The encoded VA sequence is then projected into the affective feature $f_{E}$.

\subsection{Two-Stage Motion Generation}
\label{sec_3_3}
\paragraph{\textbf{Deterministic Motion Prior (DMP)}}
As shown in Figure~\ref{fig:pipeline1}, the first stage predicts a motion prior for subsequent residual refinement. To prevent inaccurate priors from compromising residual learning, DMP is constrained in two ways: (a) Input: It relies solely on the frozen WavLM content feature $f_A^{cont}$, shape parameter $\beta$, and previous motion context $M_p$, bypassing the weak-control signals from Sec.~\ref{sec_3_2}. (b) Output: It estimates only a compact target $\tilde{M} = [\psi, \theta_{jaw}]$, where $\theta_{jaw}$ represents the jaw-opening parameter. Deferring global head rotation simplifies the first-stage prediction, leaving a much cleaner residual target for subsequent modeling.
Concretely, DMP takes the shape parameter $\beta$, previous motion context $M_p$, and a set of learnable queries $q_M \in \mathbb{R}^{T_c \times d}$ as inputs. Inspired by query-based architectures~\cite{Q-Former}, $q_M$ serves as latent placeholders for the current segment. The model maps $M_p$ and $\beta$ into latent embeddings and applies audio-conditioned Transformer decoding using the frozen speech feature $f_A^{cont}$. Finally, a motion decoder outputs the compact prior $\tilde{M}^{prior}$, which is restored to the full motion space by padding a normalized zero head pose: $M_c^{prior}=[\psi^{prior}, \mathbf{0}, \theta_{jaw}^{prior}]$.
To ensure spatial accuracy and temporal consistency, we supervise DMP in the FLAME vertex space.
Formally, let $\mathbf{V}$ and $\hat{\mathbf{V}}$ denote the target and predicted FLAME vertices, respectively, with the labels $lip$ and $face$ specifying the corresponding regions. We then define geometric losses in the FLAME vertex space: vertex loss $\mathcal{L}_\mathrm{vert}$ to enforce spatial reconstruction accuracy, and velocity loss $\mathcal{L}_\mathrm{vel}$ to improve temporal consistency:
\begin{equation}
\mathcal{L}_\mathrm{vert} = 
\lambda_{vert}^{face} \left\|\hat{\mathbf{V}}_\mathrm{face} - \mathbf{V}_\mathrm{face}\right\|_1 + 
\lambda_{vert}^{lip} \left\|\hat{\mathbf{V}}_\mathrm{lip} - \mathbf{V}_\mathrm{lip}\right\|_1,
\end{equation}
\begin{equation}
\mathcal{L}_\mathrm{vel} = 
\lambda_{vel}^{face}\left\|\Delta \hat{\mathbf{V}}_\mathrm{face} - \Delta \mathbf{V}_\mathrm{face}\right\|_2^2 +
\lambda_{vel}^{lip}\left\|\Delta \hat{\mathbf{V}}_\mathrm{lip} - \Delta \mathbf{V}_\mathrm{lip}\right\|_2^2,
\end{equation}
where $\Delta$ denotes the first-order temporal difference. The final objective of DMP is:

\begin{equation}
\mathcal{L}_{\mathrm{DMP}}
= \mathcal{L}_\mathrm{vert} + \mathcal{L}_\mathrm{vel}
\end{equation}

\paragraph{\textbf{Residual Flow Matching (RFM)}}
Given motion prior $M_c^{prior}$, the second stage RFM only model the residual motion $M_c^{res} = M_c - M_c^{prior}$. To alleviate scale variation across motion channels, we normalize $M_c^{res}$ and use it as the target endpoint $x_1$, thus casting flow matching in the residual space. We sample $x_0 \sim \mathcal{N}(0, I)$ and define the interpolation path as
\begin{equation}
x_{t'} = (1-t')x_0 + t'x_1.
\end{equation}
The velocity network $V_\theta(x_{t'}, t', c)$ predicts the vector field from the noisy state $x_{t'}$, conditioned on the deterministic prior $M_c^{prior}$ and $c = \{f_A, f_E, f_G\}$. Here, $f_A$ and $f_E$ are injected via Multi-Head Cross-Attention, while global feature $f_G$ are injected through AdaLN~\cite{DiT}. In addition, RoPE~\cite{RoPE} is used in self-attention, and local causal masks are adopted in cross-attention to preserve temporal order and improve alignment.

Given the conditioning set $c$, the model is trained with the standard flow-matching objective:
\begin{equation}
\mathcal{L}_\mathrm{FM} = \mathbb{E}_{x_0,x_1,t'} \left[ \left\|V_\theta(x_{t'}, t', c) - (x_1 - x_0)\right\|_2^2 \right].
\end{equation}

Furthermore, rather than imposing a deterministic endpoint regression that compromises stochasticity, we explicitly regularize the reconstructed motion $\hat{M}_c = M_c^{\mathrm{prior}} + \hat{M}_c^{res}$ using velocity loss $\mathcal{L}_\mathrm{vel}$, smoothness loss $\mathcal{L}_\mathrm{smooth}$, and standard-deviation loss $\mathcal{L}_\mathrm{std}$. 
Let $\mathbf{S}_r$ and $\hat{\mathbf{S}}_r$ denote the target and predicted trajectories for facial region $r\in\{\text{face, lips, brows, eyes}\}$ or head-pose axis $r\in\{\text{head}\}$, respectively. Three geometric auxiliary losses are defined as:
\begin{equation}
\mathcal{L}_\mathrm{vel} = \sum_{r \in \mathcal{R}_1} \lambda_r^{vel} \left\| \Delta \hat{\mathbf{S}}_r - \Delta \mathbf{S}_r \right\|_2^2,
\end{equation}
\begin{equation}
\mathcal{L}_\mathrm{smooth} = \sum_{r \in \mathcal{R}_2}\lambda_r^{smooth} \left\| \Delta \hat{\mathbf{S}}_r^{\,t} - \Delta \hat{\mathbf{S}}_r^{\,t-1} \right\|_2^2,
\end{equation}
\begin{equation}
\mathcal{L}_\mathrm{std} = \sum_{r \in \mathcal{R}_3} \lambda_r^{std} \left\| \operatorname{std_t}(\hat{\mathbf{S}}_r) - \operatorname{std_t}(\mathbf{S}_r) \right\|_2^2,
\end{equation}
where $\mathcal{R}_1=\{\text{face, lips, brows, eyes, head}\}$, $\mathcal{R}_2=\{\text{face, head}\}$, and $\mathcal{R}_3=\{\text{brows, eyes, head}\}$. The detailed formulations of these auxiliary geometric losses are provided in the supplementary material.

The final objective of RFM is:
\begin{equation}
\mathcal{L}_\mathrm{RFM} = \mathcal{L}_\mathrm{FM} + \mathcal{L}_\mathrm{vel} + \mathcal{L}_\mathrm{smooth} + \mathcal{L}_\mathrm{std}.
\end{equation}
\begin{figure}[t]
  \centering
  \includegraphics[width=0.85\columnwidth]{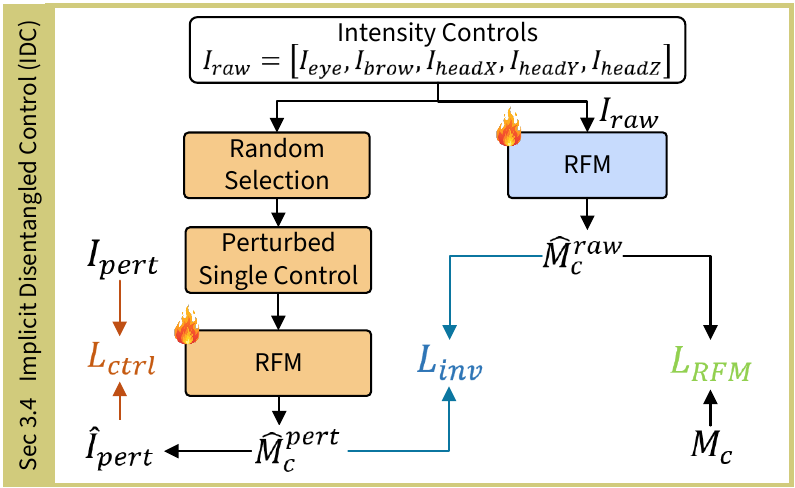}
\caption{Illustration of the implicit disentangled control (IDC) strategy (Sec.~\ref{sec_3_4}). At each step, only one control dimension in $I_{\mathrm{raw}}$ is randomly selected and perturbed to form $I_{\mathrm{pert}}$. This drives the intended motion change via $L_{\mathrm{ctrl}}$, while $L_{\mathrm{inv}}$ preserves the remaining controls.}
  \Description{Pipeline of SubtleTalk(img2).}
  \label{fig:pipeline2}
\end{figure}
\subsection{Robust and Adaptive Control}
\label{sec_3_4}

\paragraph{\textbf{Implicit Disentangled Control (IDC)}}
Although regional intensity signals provide explicit amplitude control, directly injecting them into the generator may still cause cross-factor interference, especially when one control factor is adjusted while the remaining factors are expected to stay unchanged. To alleviate this issue, we introduce an implicit disentangled control strategy (IDC) after an initial warm-up of second-stage training, when the residual generator RFM has learned a sufficiently stable base mapping.

Let 
$I_{raw} = [I_{eye}, I_{brow}, I_{headX}, I_{headY}, I_{headZ}]$
denote the original intensity controls. 
As shown in Figure \ref{fig:pipeline2}, at each iteration, we randomly select a control index $k$ and perturb only that dimension by scaling it with a random factor $\alpha$, i.e.,
\begin{equation}
I_{pert}^{\,j} =
\begin{cases}
\alpha I_{raw}^{\,j}, & j=k,\\
I_{raw}^{\,j}, & j\neq k,
\end{cases}
\end{equation}

A straightforward alternative is to apply the perturbation in a single forward pass and impose the control constraints directly on the resulting output. However, this would entangle perturbation-driven regularization with the learning of the base speech-to-motion mapping in the same branch, which is less favorable for stable optimization. We therefore adopt a dual-branch formulation. 
Specifically, the model utilizes shared inputs, parameters, and generative states to produce a raw prediction $\hat{M}_c^{raw}$ and a perturbed prediction $\hat{M}_c^{pert}$:
\begin{equation}
\hat{M}_c^{raw} = \mathrm{RFM}(A_w, M_p, f_A, f_E, \beta, I_{raw})
\end{equation}
\begin{equation}
\hat{M}_c^{pert} = \mathrm{RFM}(A_w, M_p, f_A, f_E, \beta, I_{pert}).
\end{equation}

To achieve disentanglement, we apply two complementary constraints on the perturbed branch, using the raw branch as a fixed reference. 
First, to ensure the targeted factor faithfully follows the perturbed magnitude, we define a control loss $\mathcal{L}_{\mathrm{ctrl}}$:
\begin{equation}
\mathcal{L}_{\mathrm{ctrl}} = \left\|\hat{I}_{pert}^{\,k} - I_{pert}^{\,k}\right\|_2^2,
\end{equation}
Second, to prevent the perturbation from leaking into unrelated dimensions, we impose an invariance loss $\mathcal{L}_{\mathrm{inv}}$ on the complementary components:

\begin{equation}
\mathcal{L}_{\mathrm{inv}} =
\left\|
\mathrm{sg}\!\left[\mathrm{\hat{S}}_{\neg k}^{raw}\right]
-
\mathrm{\hat{S}}_{\neg k}^{pert}
\right\|_2^2,
\end{equation}
where $\mathrm{\hat{S}}_{\neg k}$ denotes the motion components excluding the perturbed factor $k$, and $\mathrm{sg}[\cdot]$ denotes the stop-gradient operation. By detaching the raw branch, the refinement losses act only on the perturbed branch, while the raw branch remains directly supervised by the original second-stage objective $\mathcal{L}_{\mathrm{RFM}}$.

The final objective of IDC is:
\begin{equation}
\mathcal{L}_{\mathrm{IDC}} = \mathcal{L}_{\mathrm{RFM}} + \lambda_{ctrl}\mathcal{L}_{\mathrm{ctrl}} + \lambda_{inv}\mathcal{L}_{\mathrm{inv}}.
\end{equation}

\paragraph{\textbf{VA Dynamics Predictor (VADP)}}
During training, SubtleTalk is conditioned on dense frame-wise VA signals, which are typically unavailable at inference time. We therefore introduce a lightweight VA Dynamics Predictor (VADP) to support both automatic and user-guided affect control. Given a speech window $A_w$, VADP extracts speech-affect features from a frozen emotion2vec~\cite{emotion2vec} encoder followed by temporal convolutions. When the user provides sparse VA anchors on a frame set $\Omega$, they are encoded and injected through feature modulation; otherwise, the prediction relies solely on speech. The network predicts a dense VA trajectory $\overline{VA}_{pred}$, and the final control signal is obtained by enforcing the user-specified anchors:
\begin{equation}
VA_{pred}(t)=
\begin{cases}
VA_{anchor}(t), & t \in \Omega,\\
\overline{VA}_{pred}(t), & t \notin \Omega.
\end{cases}
\end{equation}

\section{Experiments}
\subsection{Datasets}
We largely follow the data construction pipeline of TFHP~\cite{DiffPoseTalk}, while focusing on three aspects: larger-scale data collection, more rigorous data curation, and higher-quality pseudo-labeling using a stronger reconstruction model.
First, we collect source videos from five datasets: HDTF~\cite{HDTF}, CelebV-HQ~\cite{CelebV-HQ}, VFHQ~\cite{VFHQ}, and TFHP~\cite{DiffPoseTalk}, which are in-the-wild 2D audio-visual datasets, together with MEAD~\cite{MEAD}, a dataset collected in a controlled lab setting, to enrich emotional diversity. And all videos are resampled to 25 fps and all audio tracks to 16 kHz.
Second, to ensure high-quality pseudo-labels, we apply a rigorous filtering pipeline, including removing clips with poor audio-visual synchronization, severe visual corruption, or extreme head poses. 
Specifically, for audio-visual synchronization, we use SyncNet~\cite{SyncNet} to filter out clips with poor lip synchronization. 
For head-pose filtering, we mainly control large yaw angles, i.e., left-right head turns, and discard any clip in which the yaw exceeds $40^\circ$ in more than 10\% of its frames, since excessive side views often cause self-occlusion and unstable pseudo-labeling.
Third, we estimate full FLAME parameters, including shape, expression, and pose, using the state-of-the-art monocular 3D reconstruction method TEASER~\cite{TEASER}, which provides higher-quality reconstruction of upper-face motions. 
We also apply a Savitzky-Golay filter~\cite{SG-Filter} to the expression and pose sequences to reduce temporal jitter.
In addition, we extract frame-wise valence-arousal (VA) labels using EmotiEffLib~\cite{EmotiEffLib}, providing continuous affect supervision aligned with the motion sequences.

As summarized in Table~\ref{tab:subtletalkface_stats}, the final dataset contains 36,733 clips (73.83 hours) from 3,905 identities.
We divide the dataset into training, validation, and test sets with strictly disjoint identities within each source dataset.
Compared with existing 3D facial animation datasets (Table~\ref{tab:dataset_comparison}), \textbf{SubtleTalk-Face} provides a stronger balance of scale, identity diversity, and in-the-wild coverage, supported by stricter quality control and higher-quality 3D annotations.

\begin{table}
\caption{Statistics of \textbf{SubtleTalk-Face}, including the selected subsets from each source dataset and the identity-disjoint train/validation/test split.}
  \label{tab:subtletalkface_stats}
  \centering
  \begin{tabular}{lccc}
    \toprule
    Source & Clips & Identities & Duration (h) \\
    \midrule
    MEAD~\cite{MEAD} & 29,533 & 45 & 34.59 \\
    HDTF~\cite{HDTF}        & 381    & 322 & 14.80 \\
    CelebV-HQ~\cite{CelebV-HQ}   & 3,060  & 1388 & 7.75 \\
    VFHQ~\cite{VFHQ}        & 3,136  & 1848 & 7.51 \\
    TFHP~\cite{DiffPoseTalk}        & 623    & 302 & 9.18 \\
    \midrule
    \textbf{Total}       & \textbf{36,733} & \textbf{3,905} & \textbf{73.83} \\
    \midrule
    Train       & 29,578 & 2,456 & 59.76 \\
    Validation         & 3,854  & 724   & 7.40 \\
    Test        & 3,301  & 725   & 6.67 \\
    \bottomrule
  \end{tabular}
\end{table}

\begin{table}
\caption{Comparison with representative 3D facial animation datasets. ``Lang.'' denotes the primary language setting, and ``Env.'' denotes the acquisition environment.}
  \label{tab:dataset_comparison}
  \centering
  \begin{tabular}{lcccc}
    \toprule
    Dataset & Identities & Duration (h) & Lang. & Env.\\
    \midrule
    BIWI~\cite{BIWI} & 14 & 1.44 & EN & Lab \\
    VOCASET~\cite{VOCA} & 12 & 0.5 & EN & Lab \\
    MeshTalk~\cite{MeshTalk} & 250 & 13 & EN & Lab \\
    3D-ETF~\cite{EmoTalk} & 100+ & 6.5 & EN & Mix \\
    MEAD-3D~\cite{EMOTE} & 60 & 38 & EN & Lab \\
    TFHP~\cite{DiffPoseTalk} & 588 & 26.5 & EN & Mix \\
    MMHead~\cite{MMHead} & 2,000+ & 49 & Mul. & Mix \\
    \textbf{SubtleTalk-Face} & \textbf{3,905} & \textbf{73.83} & \textbf{Mul.} & \textbf{Mix} \\
    \bottomrule
  \end{tabular}
\end{table}

\subsection{Experimental Settings}
\paragraph{\textbf{Implementation Details}}
Our training pipeline consists of two stages. In Stage 1, we train the DMP module using the loss $L_{\mathrm{DMP}}$. In Stage 2, we optimize the RFM module in a two-phase manner: we first perform a warm-up stage using only $L_{\mathrm{RFM}}$, and then activate the IDC strategy in the later stage to further encourage disentangled optimization with $L_{\mathrm{IDC}}$. 
For the network architecture, DMP is built with 4 Transformer decoder blocks and 8-head multi-head cross-attention (MHCA), with a hidden dimension of 256. RFM uses 6 Transformer decoder blocks and 8-head MHCA, with the hidden dimension increased to 512. The temporal window settings are $T_p=10$, $T_c=100$, and $T_w=T_p+T_c=110$. All models are implemented in PyTorch Lightning. DMP is trained on 4 NVIDIA V100 GPUs using the AdamW~\cite{AdamW} optimizer with a learning rate of $10^{-4}$, taking approximately 2 hours. RFM is trained on the same 4 NVIDIA V100 GPUs with AdamW and a learning rate of $10^{-3}$, requiring about 6 hours for the warm-up phase and 2 hours for the IDC phase.

\paragraph{\textbf{Baseline Methods}}
We compare our method against several state-of-the-art 3D facial animation approaches, including FaceFormer~\cite{FaceFormer}, 
FaceDiffuser~\cite{FaceDiffuser}, DiffPoseTalk~\cite{DiffPoseTalk}, and ARTalk~\cite{ARTalk}. For a fair comparison, we train our model and retrain all open-source baselines on the training split of the newly constructed 73.83-hour \textbf{SubtleTalk-Face} dataset. All quantitative evaluations are performed on the corresponding test split.
\subsection{Quantitative Evaluation}

We evaluate the generated facial motion using three metrics: lip vertex error (LVE) for lip synchronization, upper face dynamics deviation (FDD) for upper-face motion dynamics, and head dynamics deviation (HDD) for head-motion dynamics. 
Detailed computation formulas for these metrics are provided in the supplementary material.
For presentation clarity, LVE in Table~\ref{tab:main_comparison} are reported in units of $10^{-4}$, while FDD and HDD are reported in units of $10^{-3}$. 

We present the quantitative results in Table~\ref{tab:main_comparison}. Our method demonstrates a significant advantage in generating weakly correlated facial dynamics, achieving remarkably lower FDD and HDD compared to all baselines. This indicates our model's superior capability in capturing subtle upper-face expressiveness and synthesizing natural, plausible head movements. Furthermore, while excelling in overall facial dynamics, our approach maintains highly competitive lip synchronization, yielding a comparable or slightly better LVE than existing methods. Overall, the evaluation confirms that our method successfully generates highly dynamic and realistic talking faces while preserving accurate speech-motion alignment.

\begin{table}
\caption{Quantitative comparison with baseline methods on the \textbf{SubtleTalk-Face} test split. Lower is better for all metrics. N/A in HDD indicates that the corresponding method does not support head motion prediction.}
  \label{tab:main_comparison}
  \centering
  \begin{tabular}{lcccc}
    \toprule
    Methods & FDD$\downarrow$ & HDD$\downarrow$ & LVE$\downarrow$ \\
    \midrule
    FaceFormer~\cite{FaceFormer} & 15.39 & N/A &  13.87 \\
    FaceDiffuser~\cite{FaceDiffuser} & 14.54  & N/A & 14.30 \\
    DiffPoseTalk~\cite{DiffPoseTalk} & 9.74 & 21.71 &  13.72 \\
    ARTalk~\cite{ARTalk} & 11.37 & 26.70 & 11.98 \\
    \textbf{Ours} & \textbf{4.46} & \textbf{6.61} &  \textbf{11.96} \\
    \bottomrule
  \end{tabular}
\end{table}

\subsection{Qualitative Evaluation}

\begin{figure*}[t]
  \centering
  \includegraphics[width=0.85\textwidth]{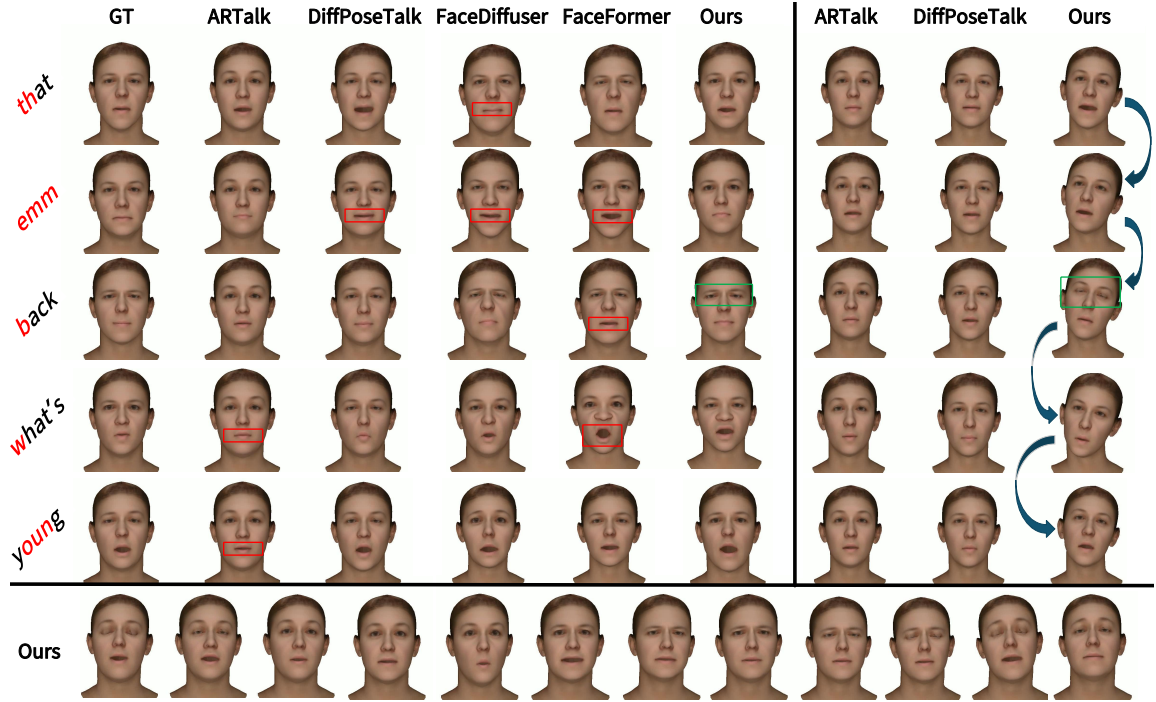}
\caption{
Qualitative comparison with baseline methods. We contrast lip shapes (left) and head motions (right) against baselines. The bottom row displays a continuous sequence demonstrating our method's dynamic eyebrow and eye motions over time.
}
  \Description{Qualitative comparison}
  \label{fig:exp1}
\end{figure*}

For clearer qualitative visualization, we adopt two rendering strategies. First, to highlight subtle facial deformations, we apply a fixed FLAME texture, acknowledging \cite{FLAME, StyleGAN, PhotoOpt}. Second, following MMHead~\cite{MMHead}, we apply the predicted head pose to the neck joint to prevent rigid torso rotation. 
In Figure~\ref{fig:exp1}, we show qualitative comparisons of our method with baseline methods.
Our method can generate natural upper-face behaviors, especially in the eyebrow and eye regions, and 
produces head movements with more plausible motion amplitudes. 
Meanwhile, our method still preserves stable lip motions that remain well aligned with the input speech.

\subsection{User Study}

\begin{table}
\caption{User study results. Percentages represent the proportion of participants preferring our method over the baselines. Face Nat. and Head Nat. denote facial and head naturalness, respectively. Lip Sync evaluates audio-lip alignment.}
  \label{tab:user_study}
  \centering
  \begin{tabular}{lccc}
    \toprule
    Methods & Face Nat. & Head Nat. & Lip Sync \\
    \midrule
    vs DiffPoseTalk~\cite{DiffPoseTalk} & 94.81\%  & 74.44\% & 69.63\% \\
    vs ARTalk~\cite{ARTalk} & 86.30\% & 78.89\% & 51.48\% \\
    \bottomrule
  \end{tabular}
\end{table}

To evaluate performance in real-world application scenarios, we conduct a subjective user study with 27 participants. We source 10 out-of-domain audio clips from the internet, ranging from 5 to 30 seconds and covering diverse emotions (angry, fear, happy, sad, surprise, and neutral). Using solely these audio clips as input, we generate animation videos for pairwise comparison. In each trial, the animation generated by our method is displayed side-by-side with a baseline method in a spatially randomized order (i.e., left/right positions were not fixed). Participants then choose the better result based on lip synchronization, facial naturalness, and head naturalness. As summarized in Table~\ref{tab:user_study}, the strong user preference in facial and head naturalness suggests that our approach makes progress in synthesizing weakly speech-correlated dynamics.

\begin{table}
\caption{Ablation study of the proposed method on the \textbf{SubtleTalk-Face} test split. N/A in HDD indicates that the corresponding method does
not predict head pose.}
  \label{tab:ablation}
  \centering
  \begin{tabular}{lccc}
    \toprule
    Methods & FDD$\downarrow$ & HDD$\downarrow$ & LVE$\downarrow$ \\
    \midrule
    DMP (only) & 15.25 & N/A & \textbf{11.35} \\
    FM (only) & 12.47 & 27.34 & 14.60 \\
    FM+DMP & 11.44 & 25.57 &  12.46 \\
    FM+MultiCond & 8.02 & 12.58 & 15.05 \\
    FM+MultiCond+IDC & 5.78 & 7.93 & 14.51  \\
    FM+DMP+MultiCond & 5.21 & 6.96 & 12.59 \\
    \textbf{Full model} & \textbf{4.46} & \textbf{6.61} & 11.96 \\
    \bottomrule
  \end{tabular}
\end{table}

\subsection{Ablation Study}
We conduct ablation studies to assess the contribution of each key component to modeling weakly correlated facial dynamics (Table~\ref{tab:ablation}):
\textbf{(a) DMP-only.} Using only DMP achieves the best LVE but performs poorly on FDD. This reflects the tendency of regression-based modeling to favor smooth averaged solutions, which are inadequate for capturing multimodal weakly correlated dynamics.
\textbf{(b) FM vs. FM+DMP.} DMP mainly reduces LVE, while its effect on FDD/HDD is limited, suggesting that it stabilizes strongly audio-correlated mouth motion.
\textbf{(c) FM vs. FM+MultiCond vs. FM+MultiCond+IDC.} Introducing multiple conditioning signals (VA, intensity, and prosody) reduces FDD and HDD by controlling weakly audio-correlated factors such as upper-face and head motion, confirming that audio alone is insufficient to model upper-face and head dynamics. However, without IDC, different control signals may become entangled, slightly increasing LVE. IDC helps improve control decoupling, thereby better balancing lip accuracy with upper-face and head-motion dynamics.
\textbf{(d) FM+MultiCond vs. FM+DMP+MultiCond.} DMP further improves all metrics, consistent with our motivation that DMP stabilizes strongly audio-correlated mouth motion and FM focuses on residual motion.
\textbf{(e) Full model.} By integrating FM, DMP, MultiCond, and IDC, the full model combines the complementary advantages of all components, achieving the best FDD and HDD while maintaining competitive LVE, thereby providing the best overall balance between lip accuracy and weakly correlated upper-face and head-motion dynamics.

\section{Conclusion}
In this paper, we introduce SubtleTalk, a two-stage residual flow-matching framework for generating natural and controllable weakly correlated 3D facial dynamics from speech. By integrating explicit multi-condition modeling of motion intensity, prosody and valence-arousal, SubtleTalk enables fine-grained controllability and alleviates mean-collapse to accurately capture subtle facial events. To mitigate current data limitations, we further construct SubtleTalk-Face, a large-scale, high-fidelity dataset for 3D facial animation. Extensive experiments demonstrate that SubtleTalk significantly outperforms existing methods in generating diverse, realistic, and speech-consistent animations. 

\begin{acks}
This work was supported by the National Natural Science Foundation of China (NSFC, No. 62472285 and No. 62102255) and Tencent Industry-University Collaboration Research Funding.
\end{acks}

\bibliographystyle{ACM-Reference-Format}
\balance
\bibliography{SubtleTalkRef}

\end{document}